\RequirePackage{ifpdf}
\documentclass[12pt,letterpaper]{JHEP3}         
\usepackage{amssymb,amsfonts}

\usepackage{wrapfig}
\usepackage{epsfig}

\def\be{\begin{eqnarray}}
\def\ee{\end{eqnarray}}
\newcommand{\nn}{\nonumber}
\newcommand\para{\paragraph{}}

\newcommand{\eqn}[1]{(\ref{#1})}

\def\Dslash{\,\,{\raise.15ex\hbox{/}\mkern-12mu D}}
\def\Dbarslash{\,\,{\raise.15ex\hbox{/}\mkern-12mu {\bar D}}}
\def\delslash{\,\,{\raise.15ex\hbox{/}\mkern-9mu \partial}}
\def\delbarslash{\,\,{\raise.15ex\hbox{/}\mkern-9mu {\bar\partial}}}
\def\pslash{\,\,{\raise.15ex\hbox{/}\mkern-9mu p}}
\def\calDslash{\,\,{\raise.15ex\hbox{/}\mkern-12mu {\cal D}}}

\newcommand{\sign}{{\rm sign}}

\def\lae{\mathrel{\mathop{\smash{\lower .5 ex \hbox{$\stackrel<\sim$}}}}}
\def\lae{\mathrel{\mathop{\smash{\lower .5 ex \hbox{$\stackrel>\sim$}}}}}

\title{Particle-Vortex Duality from 3d Bosonization}

\author{Andreas Karch${}^1$ and David Tong${}^{2,3}$\\
${}^1$Department of Physics, \\
University of Washington, Seattle, WA 98195, USA \\
${}^2$Department of Applied Mathematics and Theoretical Physics, \\
University of Cambridge, Cambridge, CB3 OWA, UK \\
${}^3$Stanford Institute for Theoretical Physics \\
Via Pueblo, Stanford, CA 94305, USA\\
{\tt  akarch@uw.edu, d.tong@damtp.cam.ac.uk}\\
}

\abstract{We provide a simple derivation of particle-vortex duality in $d=2+1$ dimensions. Our starting point is a relativistic form of
flux attachment, designed to  transmute the statistics of particles. From this seed,  we derive a web of new dualities. These include
 particle-vortex duality for bosons as well as the recently discovered counterpart for fermions.}

\begin{document}
\pagestyle{plain} \setcounter{page}{1}
\newcounter{bean}
\baselineskip16pt \setcounter{section}{0}

\section{Introduction: A Collection of Dualities}

Many quantum field  theories in $d=2+1$ dimensions enjoy a beautiful property known as particle-vortex duality. This relates two different theories, with the fundamental fields of one theory mapped to vortices -- or, more precisely, monopole operators --  of the other. The duality has proven to be a powerful tool in a number of different settings, ranging from condensed matter physics to string theory.

\para
Particle-vortex duality for bosonic systems was established long ago \cite{peskin,dh}. In the simplest version, the duality relates the theory of a complex scalar field (the XY model)
\be S = \int d^3x\ |(\partial_\mu - iA_\mu)\phi|^2 - V(\phi)\label{xy}\ee
to the Abelian-Higgs model
\be S = \int d^3x\ |(\partial_\mu - ia_\mu)\Phi|^2 - \tilde{V}(\Phi) + \frac{1}{2\pi}\epsilon^{\mu\nu\rho}A_\mu \partial_\nu a_\rho\label{ah}\ee
In the actions above, $A_\mu$ is a background gauge field. Its coupling to the currents  in the two theories reveals that the particle density of $\phi$ in \eqn{xy}  is equated to the flux density $f/2\pi=da/2\pi$ in \eqn{ah}. This is the essence of particle-vortex duality.

\para
More recently, an analogous  duality was  proposed for fermions. The free Dirac fermion with action
\be S = \int d^3x\ i\bar{\psi}\gamma^\mu(\partial_\mu - iA_\mu)\psi\label{son1}\ee
is conjectured to be dual to QED${}_3$ with a single species of fermion
\be S  = \int d^3x\ i\bar{\Psi}\gamma^\mu(\partial_\mu - ia_\mu)\Psi  +\frac{1}{4\pi}\epsilon^{\mu\nu\rho} A_\mu \partial_\nu a_\rho\label{son2}\ee
This is more subtle, not least because a single Dirac fermion in $d=2+1$ dimensions flirts with the parity anomaly. This is avoided in the above theories by changing the flux quantisation conditions of the gauge field; ultimately, this seems to be sensible only when the  theories are viewed as living on the boundary of a $d=3+1$ dimensional system.

\para
The proposed fermionic particle-vortex duality lies at the heart of a number of inter-related topics in condensed matter physics. The duality  first arose  in Son's suggestion that the correct description of the half-filled Landau level involves an emergent (``composite") Dirac fermion \cite{son}. Shortly afterwards, it was realised that the duality plays an important role in describing the surface states of interacting  topological insulators \cite{senthil,ashvin}.
These ideas have subsequently been extended in a number of different directions \cite{senthil2,ashvin2,senthil3,mike,max}, including a derivation of the duality starting from an array of $d=1+1$ dimensional wires \cite{mross}.

\subsection*{Flux Attachment}

There is a second, seemingly unrelated, operation that one can  perform in $d=2+1$ dimensions. This is statistical transmutation of particles through flux attachment \cite{frank}. Attaching a single quantum of flux to a particle turns a boson into a fermion and vice versa, while adding  two quanta of flux leaves the statistics unchanged. This process underlies the original concept of composite fermions as emergent particles in the lowest Landau level \cite{jjain}.

\para
The idea of flux attachment was first applied to non-relativistic particles. However, there also exist versions of flux attachment for relativistic particles. These take the form of dualities between bosonic and fermionic field theories, where one side is coupled to a Chern-Simons interaction to implement the statistical transmutation. See, for example, \cite{polyakov,shankar1,shankar2,fradhap,cfw,john}. These dualities are sometimes referred to as {\it 3d bosonization} and will be the starting point of the present paper.

\para
As an aside, we note that there has recently been a revival of this idea in the context of non-Abelian $U(N)$  gauge theories \cite{shiraz,guy1,guy2,ofer}, where the $1/N$ expansion allows a number of very impressive checks of the duality \cite{shiraz2,karthik,shiraz3,sean}. This too leads to a compelling story for finite $N$ and, ultimately, for the $U(1)$ gauge theories of interest in this paper.  Combining studies of  supersymmetric theories  \cite{givkut,benini,ofer3d4d,parksandrecreation}, RG flows \cite{jain,guy3}, operator maps \cite{radicevic} and level-rank dualities \cite{levelrank} results in a consistent picture. These different threads were tied together in a careful analysis  by  Aharony \cite{ofer}.

\subsubsection*{Synthesis}

In this paper, we will need only the Abelian version of the 3d bosonization duality. This is the simplest relativistic generalisation of the standard flux attachment story: a scalar coupled to a  $U(1)$ Chern-Simons term at level 1 is equivalent to a fermion. We will describe this in more detail in Section \ref{bosonsec}.

\para
From this conjectured bosonization duality, all else follows. This is the topic of Section \ref{pvsec}.  By manipulating the path integral, we derive a web of further dualities. These include both the bosonic and fermionic versions of particle-vortex dualities presented above, as well as many more. All these derivations hold at the level of the partition function, viewed here as a function of  background gauge fields (i.e. $A_\mu$ in the examples above). This means that all correlation functions of currents agree on both sides.

\subsubsection*{Supersymmetry}

There is one further ingredient that we would like to add to the mix. This is the supersymmetric version of particle-vortex duality, known also as {\it mirror symmetry}. First discovered in  \cite{intseib}, there are now many examples which differ in their gauge symmetry, matter content and amount of supersymmetry; see \cite{horin4,horin2,ahiss,doreytong,me} for a small sample. Mirror symmetry was applied to the problem of the half-filled Landau level in \cite{shamit}, although the existence of a pair of Dirac fermions means that the resulting physics is rather different from the fermionic duality of \eqn{son1} and \eqn{son2}.

\para
The power of supersymmetry provides greater control over the strong coupling regions of these theories. This means that one can be much more precise about the operator map between theories and, in certain cases, one can effectively prove  the duality by showing that the monopole operator is indeed a free field \cite{kapmono}. Particularly pertinent for the present paper are the path integral manipulations developed in \cite{kapstrass} for supersymmetric theories. We will borrow heavily from these ideas.

\para
Since the supersymmetric mirror pairs involve both bosons and fermions, one might imagine that they can be constructed by combing the bosonic and fermionic dualities described in \eqn{xy}-\eqn{son2}. However, the way the parity anomaly plays out in the supersymmetric theories is rather different from the way it works in the purely fermionic duality of \cite{son,senthil,ashvin}. Instead, following the large $N$ analysis of \cite{jain,guy3}, we suggest that one should think of mirror symmetry more as a bosonization, with bosonic currents in one theory mapped to fermionic currents in the other. We discuss this in Section \ref{dissec}.

\para
{\bf Note Added:} Before submission to the Arxiv, we became aware of \cite{ssww} which contains overlap with this paper. Related work is also contained in \cite{mn}.

\section{Bosonisation in $d=2+1$}\label{bosonsec}

We start in this section by describing the relativistic generalisations of flux attachment that we will need \cite{polyakov,fradhap,cfw,john,ofer}.

\subsection{Building Blocks}

Throughout this paper, we will work with three types of fields: complex scalars $\phi$, two-component Dirac spinors $\psi$ and Abelian gauge fields. The latter fall into two categories: background gauge fields, which we will initially denote as $A_\mu$  and dynamical gauge fields $a_\mu$. In the condensed matter context, $A_\mu$ is usually to be thought of as electromagnetism while $a_\mu$ is an emergent gauge field.

\para
All gauge fields, whether background or dynamical, are compact in the sense that the fluxes are quantised. It will be somewhat easier to discuss this flux quantisation if we take our spatial slices to be ${\bf S}^2$, rather than ${\bf R}^2$,
The precise choice of the quantisation condition will be an important part of the story and we will specify it afresh for each theory. For now, we recall the standard story of Dirac quantisation: if fundamental fields have unit charge, then the flux is quantised as
\be \int_{\bf S^2} \frac{F}{2\pi} \in {\bf Z}\label{diracq}\ee
where $F=dA$.

\para
We will insist that our partition functions are gauge invariant. Of course, this has to be the case for the dynamical gauge fields $a$; however, we will also insist that our partition functions are gauge invariant for the background gauge field $A$. This is particularly relevant in the presence of Chern-Simons terms\footnote{For all gauge fields, we only write the Chern-Simons terms explicitly. For dynamical gauge fields, there is also an implicit Maxwell term $\frac{1}{g^2}f_{\mu\nu}f^{\mu\nu}$. We neglect this as we are ultimately interested in the infra-red limit $g^2\rightarrow \infty$. Nonetheless, we should remember that in the presence of an ultra-violet cut-off $\Lambda_{UV}$, we keep  $g^2\ll \Lambda_{UV}$ as this limit is taken.}
\be  S_{CS}[A] = \frac{1}{4\pi}\int d^3x\ \epsilon^{\mu\nu\rho}A_\mu\partial_\nu A_\rho\label{cs}\ee
This appears in the path integral as $e^{ik\,S_{CS}[A]}$ where the coefficient $k$ is referred to as the level. If $A$  obeys the standard quantisation condition \eqn{diracq} then  gauge invariance requires
\be k \in {\bf Z}\nn\ee
Since this argument is important, let us remind ourselves of the key elements. We  work on Euclidean spacetime ${\bf S}^1\times {\bf S}^2$. This allows us to introduce a new ingredient: large gauge transformations of the form $g=e^{i\theta}$, where $\theta\in [0,2\pi)$ is the coordinate of the ${\bf S}^1$. When evaluated on a flux background, the Chern-Simons action shifts under such a large gauge transformation: $\Delta S_{CS}[A] = 2\pi\int_{{\bf S}^2}{F}/{2\pi}$.
With the usual Dirac quantisation condition \eqn{diracq}, we learn that $e^{ikS_{CS}[A]}$ is gauge invariant only when $k\in {\bf Z}$ as advertised.

\para
We also need a coupling between  different Abelian gauge fields. This is achieved by a mixed Chern-Simons term, also known as a ``BF coupling",
\be S_{BF}[a;A] = \frac{1}{2\pi} \int d^3x\ \epsilon^{\mu\nu\rho} a_\mu \partial_\nu A_\rho\label{bf}\ee
The coefficient is chosen so that a flux $\int F = 2\pi$, has unit charge under $a$. The same arguments given above show that, if both $f=da$ and $F=dA$ have canonical normalisation \eqn{diracq}, then the BF-coupling must also come with integer-valued coefficient.
Note that, up to a boundary term, $S_{BF}[a;A] = S_{BF}[A;a]$.

\para
The action for the scalar fields takes the usual form
\be S_{\rm scalar}[\phi;A] = \int d^3x\ |(\partial_\mu - i A_\mu)\phi|^2 + \ldots\label{scalar}\ee
where $\ldots$ denote  possible potential terms. We will focus our attention on critical (gapless) theories. That leaves two choices: we could work with a free scalar, or we could work with a Wilson-Fisher scalar, viewed as adding a $\phi^4$ deformation and flowing to the infra-red while tuning the mass to zero. Both of these possibilities will arise below.

\para
The fermion is governed by the  Dirac action
\be S_{\rm fermion}[\psi;A] = \int d^3x\ i\bar{\psi}\gamma^\mu (\partial_\mu - iA_\mu)\psi + \ldots\label{fermion}\ee
We are interested in gapless fermions which, again, leaves two choices. One of these is  a free fermion. The other is best thought of as introducing an auxiliary field $\sigma$ and adding the term $\sigma\bar{\psi}\psi$ to the action, tuning the mass to zero. (One can play the same game to reach the Wilson-Fisher fixed point for the boson.)

\para
If $A$ is taken to obey the standard  quantisation condition \eqn{diracq} then the partition function involving the action \eqn{fermion} for a single Dirac fermion is not gauge invariant. This is the parity anomaly \cite{redlich,semenoff}.
One way to see this is  to give the fermion a  mass $m\bar{\psi}\psi$ . Integrating them out then results in the Chern-Simons term
\be \frac{1}{2}\,\sign(m) S_{CS}[A]\nn\ee
But, as described above, Chern-Simons terms are only gauge invariant with integer coefficients.

\para
Alternatively, we can see the lack of gauge invariance directly when $m=0$. Consider the background in which we insert a single unit of flux \eqn{diracq} through a spatial ${\bf S}^2$. The Dirac fermion has  a single, complex zero mode, $\chi$. This means that the monopole has two ground states,
\be |0\rangle\ \ \ {\rm and}\ \ \ \chi^\dagger |0\rangle\label{twostates}\ee
Because $\psi$ has charge $1$, the charge of these two states must differ by $+1$. But, by $CT$  symmetry  the magnitude of the charge should be the same for the two states. The net result is that we have a simple example of charge fractionalisation and the  states have charge $Q=\pm \frac{1}{2}$.
%
%
This means that, in the presence of an odd number of background fluxes, the gauge charge is not integer valued. This is in contradiction with our original Dirac quantisation condition which assumed unit fundamental charge.
Something has to break. That something is gauge invariance.

\para
The upshot of these arguments is that we must amend the action \eqn{fermion} in some way in order to preserve gauge invariance. There are (at least) two remedies. The first is to retain the quantisation condition \eqn{diracq} but include a compensating half-integer Chern-Simons action $S_{CS}[A]$. The second is to change the quantisation condition \eqn{diracq}. Both remedies will appear in different places below.

\subsection{Attaching Flux to Scalars}

With these building blocks in place, we can now describe the simple dual from which all else follows. We consider a scalar coupled to a dynamical gauge field $a$ with unit Chern-Simons coefficient. This, in turn, is coupled to a background field $A$. The full partition function takes the form
\be Z_{\rm scalar+flux}[A] = \int {\cal D}\phi {\cal D} a\ \exp\Big( iS_{\rm scalar}[\phi;a] +i S_{CS}[a] + i S_{BF}[a;A]\Big)\label{sflux}\ee
Here the path integral over gauge fields implicitly includes the relevant gauge fixing terms.
Both $f=da$ and $F=dA$ are taken to have canonical normalisation \eqn{diracq}.

\para
If we turn off the background source, so $F=0$, then the equation of motion for $a_0$ reads
\be \rho_{\rm scalar} + \frac{f}{2\pi}=0\label{scalarf}\ee
where $\rho_{\rm scalar}$ is the charge density of $\phi$. Clearly this attaches one unit of flux to each $\phi$ particle. In analogy with the familiar non-relativistic results \cite{frank}, we should expect the resulting object to be a fermion.

\para
To see this explicitly, we need to look at the monopole operator \cite{kapustin}.
 (Once again, this is simplest if we work on ${\bf S}^2$ rather than ${\bf R}^2$.) A single monopole operator has $\int f = 2\pi$. The constraint \eqn{scalarf} means that we must excite a single mode of the scalar in this background. However, the scalar monopole harmonics carry half-integer angular momentum \cite{wuyang}, ensuring that the monopole operator does indeed carry half-integer spin. The monopole is a fermion.

\para
With this in mind, we define the  fermionic path integral
\be
Z_{\rm fermion}[A]  = \int {\cal D}\psi \ \exp\Big(iS_{\rm fermion}[A] \Big)\nn\ee
As we explained previously, this is not gauge invariant. To restore gauge invariance, we dress this partition function by a Chern-Simons term for the background gauge field
with half-integer coefficient, e.g.  $e^{-\frac{i}{2}S_{CS}[A]}$. Such a term results in contact interactions between currents \cite{contact}.

\para
The proposed duality of \cite{polyakov,fradhap,ofer} is simply to identify the theory \eqn{sflux} describing scalar+flux with the fermionic theory. Their partition functions are conjectured to be related as
\be  Z_{\rm fermion}[A] \, e^{-\frac{i}{2}S_{CS}[A]}  = Z_{\rm scalar+flux}[A] \label{dual1}\ee
This is the simplest example of 3d bosonization. We note that it is also an example of a particle-vortex duality: as we saw above, the free fermion operator maps to the monopole operator in the interacting theory.

\para
In fact, the formula \eqn{dual1} actually describes two different dualities. The difference between them is hidden in the $\ldots$ in \eqn{scalar} and \eqn{fermion}. As we saw above, there are two choices for the critical scalar and fermion. The results of \cite{shiraz,guy1} strongly suggest that if we take the free fermion as the left-hand side of \eqn{dual1} then we should take the Wilson-Fisher scalar on the right-hand side. Analogously, the critical fermion with $\sigma \bar{\psi}\psi$ coupling corresponds to the free scalar.

\para
The level $-\frac{1}{2}$ of $S_{CS}[A]$ on the left-hand side of \eqn{dual1} is fixed by the Hall conductivity \cite{guy2}. To see this, let us first gap the fermion. After integrating it out, we find a Hall conductivity that is either 0 or -1 depending on the sign of the mass. On the scalar side, two different things happen depending on the sign of this fermionic mass. For one sign, the scalar is gapped and integrating out the dynamical gauge field $a$ results in a Hall conductivity $-1$; for the other sign, the scalar condenses and the gauge field $a$ is Higgsed. In this phase, the Hall conductivity vanishes. In both cases, we find agreement with the fermionic behaviour.

\para
In what follows, we will assume the duality \eqn{dual1} and use it to derive many further dualities. We do this using the kind of techniques first introduced  in \cite{kapstrass} and further explored in  \cite{witten}. First, we breathe life into the background gauge field $A$, promoting it to a dynamical gauge field. This, in turn gives rise to a new topological current ${}^\star F/2\pi$ which is subsequently coupled to a replacement background gauge field through  a BF term. Below we will use this simple but powerful trick many times. We will find that minor variations on the theme allow us to derive a vast array of different dualities, including the particle-vortex dualities described in the introduction.

\subsection{Attaching Flux to Fermions}

We start with a simple example. As described above, we promote the background gauge field $A$ in \eqn{dual1} to a dynamical field and couple it to
a new background gauge field which we denote as $C$. The left-hand side of \eqn{dual1} becomes
\be Z_{\rm fermion+flux}[C] = \int {\cal D}\psi {\cal D} A \ \exp\Big( iS_{\rm fermion}[\psi;A] - \frac{i}{2}S_{CS}[A] - i S_{BF}[A;C]\Big)\ \ \label{fermiflux}\ee
This describes a fermion coupled to a background flux. To get a feel for the resulting physics, we can again look at Gauss' law, arising as the equation of motion for $A_0$. Setting $dC=0$, this reads
\be \rho_{\rm fermion} - \frac{1}{2}\frac{F}{2\pi}=0\label{gauss1}\ee
In the background of a single monopole, $\int F=2\pi$, we must have $Q_{\rm fermion} = \frac{1}{2}$. We've already seen that this is the charge of the state $\chi^\dagger|0\rangle$ arising from quantising the zero mode \eqn{twostates}. The other state $|0\rangle$ does not satisfy Gauss' law \eqn{gauss1} and is not part of the physical Hilbert space. Moreover, the zero mode $\chi$ is known to be a singlet under rotation symmetry \cite{kapmono}. This means that the monopole operator is a boson and we might expect \eqn{fermiflux} to be dual to a scalar theory.

\para
Let us now see what becomes of the right-hand side of \eqn{dual1} under this operation. The partition function is
\be \int {\cal D}A \ Z_{\rm scalar+flux}[A]\,\exp\Big(-iS_{BF}[A;C]\Big)\nn\ee
The newly promoted gauge field $A$ appears linearly in the action and can be integrated out. Its equation of motion is simply $da=dC$. In the absence of any holonomy, we set $a=C$ to get
\be  \int D\phi\ \exp\Big(iS_{\rm scalar}[\phi;C] + i S_{CS}[C]\Big) = Z_{\rm scalar}[C] \, e^{i S_{CS}[C] }\nn\ee
This, of course, must be equal to the left-hand side \eqn{fermiflux}.
The end result is that, starting from \eqn{dual1}, we can derive a new duality in which attaching fluxes to fermions gives rise to a bosonic theory
\be Z_{\rm fermion+flux}[C] = Z_{\rm scalar}[C]  \, e^{i S_{CS}[C] }  \label{dual2}\ee
A duality of this kind first appeared in \cite{cfw} (see also \cite{john}) and has arisen more recently as a special case of non-Abelian dualities in \cite{ofer}.

\para
 One can check that repeating this procedure by gauging $C$ in \eqn{dual2} and adding a new background gauge field through a BF coupling takes us back to the duality \eqn{dual1}.

\subsection*{Time-Reversal Duals}

Before we proceed, it will be useful to highlight a  generalisation of the dualities \eqn{dual1} and \eqn{dual2}. These arise from the action of time reversal. This flips the sign of all Chern-Simons and BF couplings, leaving other terms in the action invariant. (Parity would also play the same role.) Applying time reversal to the duality \eqn{dual1} gives
\be Z_{\rm fermion}[A] \, e^{+\frac{i}{2}S_{CS}[A]} = \bar{Z}_{\rm scalar+flux}[A]\label{pdual1}\ee
where we have defined
\be \bar{Z}_{\rm scalar+flux} [A] = \int {\cal D}\phi {\cal D} a\ \exp\Big( iS_{\rm scalar}[\phi;a]  - i S_{CS}[a] - i S_{BF}[a;A]\Big)\label{pscalarflux}\ee
Similarly, applying time reversal to the duality \eqn{dual2} yields
\be \bar{Z}_{\rm fermion+flux}[C] = Z_{\rm scalar}[C]  \, e^{-i S_{CS}[C] }  \label{pdual2} \ee
where
\be \bar{Z}_{\rm fermion+flux}[C] = \int {\cal D}\psi {\cal D} A \ \exp\Big( iS_{\rm fermion}[\psi;A] + \frac{i}{2}S_{CS}[A] + i S_{BF}[A;C]\Big)\ \ \label{pfermiflux}\ee
We will have use for these versions of the duality shortly.

\section{Particle-Vortex Duality}\label{pvsec}

We can now play similar games to derive dualities which map bosons to bosons, and fermions to fermions. As we will see, these include the familiar  particle-vortex dualities.

\subsection{Bosons}

We start with the duality \eqn{dual2}. However, before we proceed, we first divide by the contact interaction so that the duality reads
\be Z_{\rm fermion+flux}[C]  \, e^{-i S_{CS}[C] }   = Z_{\rm scalar}[C]  \label{dual7}\ee
We now gauge the background field $C$. For notational reasons, it will prove useful to recycle some of our old names for gauge fields. We therefore relabel $C\rightarrow a$. We couple this to a new background gauge field which we call $A$. After gauging the right-hand side  becomes the partition function for scalar QED.
\be Z_{\rm scalar-QED}[A] =  \int {\cal D}\phi {\cal D} a \ \exp\Big( iS_{\rm scalar}[\phi; a] + iS_{BF}[a;A]\Big)\nn\ee
Now we look at the left-hand side of the duality. After these operations, the partition function is $\int {\cal D}a\ Z_{\rm fermion+flux}[a]\,e^{-iS_{CS}[a] +iS_{BF}[a;A]}$. Written out in full using \eqn{fermiflux} (and changing the names of integration variables), this reads
%
\be
\int {\cal D}\psi {\cal D}\tilde{a} {\cal D}a \ \exp\Big(iS_{\rm fermion}[\psi;\tilde{a}] - \frac{i}{2}S_{CS}[\tilde{a}] - iS_{BF}[\tilde{a};a] - iS_{CS}[a] + iS_{BF}[a;A]\Big)\nn\ee
The next step is to integrate out the gauge field $a$. Its equation of motion requires (in the absence of holonomy)  $a = A-\tilde{a}$. Substituting back in, and collecting various terms, we find the resulting partition function\footnote{This action also appears in a recent proposal for a particle-vortex symmetric description of the superconductor-insulator transition \cite{newmike}.}
%
%
%
\be
\int {\cal D}\psi {\cal D}\tilde{a}  \ \exp\Big(iS_{\rm fermion}[\psi;\tilde{a}] + \frac{i}{2}S_{CS}[\tilde{a}] - iS_{BF}[\tilde{a};A] + iS_{CS}[A] \Big)\label{midway}\ee
Something rather nice has happened: we recognise the first three terms  as the time reversed partition function $\bar{Z}_{\rm fermion+flux}$ defined in  \eqn{pfermiflux}. We can replace this by using the time reversed  duality \eqn{pdual2}. Happily, the resulting contact interaction cancels the final term in \eqn{midway}. We're left simply with the scalar partition function $Z_{\rm scalar}[A]$.

\para
We learn that applying the duality twice, once in its original form \eqn{dual2}, and once in its time reversed avatar \eqn{pdual2}, we relate two scalar partition functions
\be Z_{\rm scalar-QED}[A] = Z_{\rm scalar}[A]\nn\ee
This, of course, is the original particle-vortex duality \cite{peskin,dh}, relating  the XY model \eqn{xy} (on the right) to the Abelian Higgs model \eqn{ah} (on the left).  Following through the fate of the $\ldots$ in the original scalar action, we see that the scalar should either be free on both sides, or tuned to the Wilson-Fisher fixed point on both sides.

\para
We highlight that the derivation assumes the absence of holonomies in the gauge field when integrating out $a$. This means that the duality may be modified on ${\bf S}^2\times {\bf S}^1$, or indeed in flat space in the presence of Wilson lines.

\subsection{Fermions}

We can repeat the above derivation for the fermions. This time we start with the duality \eqn{dual1}, but only after dividing through by the contact interaction on both sides,
\be  Z_{\rm fermion}[C]   = Z_{\rm scalar+flux}[C] \, e^{+\frac{i}{2}S_{CS}[C]} \label{dual3}\ee
Now we have a problem. As we explained previously, if the background gauge field $C$ obeys the canonical quantisation condition \eqn{diracq}, then neither side of this equation is gauge invariant.

\para
A fix for this was suggested in \cite{son,senthil,ashvin}: we simply require the more stringent quantisation condition that fluxes must be even
\be \int \frac{dC}{2\pi} \in 2{\bf Z}\label{newflux}\ee
Restricted to such backgrounds, there is no anomaly.

\para
The restriction \eqn{newflux} is certainly allowed for background gauge fields which are under our control. However, the next step is to promote $C$ to a dynamical field and here the condition \eqn{newflux} is far from innocuous. A more systematic treatment of this can be found in \cite{ssww,mm}.
%

\para
Let us look at what becomes of the two sides of the duality \eqn{dual3}. The left-hand side is simply QED${}_3$, with a single flavour of fermion. Changing the name of integration variables, the partition function is
\be Z_{\rm QED}[A] = \int {\cal D}\psi{\cal D}a\ \exp\Big(iS_{\rm fermion}[\psi;a] + \frac{i}{2}S_{BF}[a;A]\Big)\label{qed}\ee
where the final term is the coupling to a background field $A$.  The partition function is gauge invariant only if $dA$ also obeys the quantisation condition \eqn{newflux}.

\para
Meanwhile, the calculation on the right-hand side follows closely the derivation of bosonic particle-vortex duality above. Only factors of 2 are different but since these factors are important, let us spell out the steps here. The partition function on the right-hand side reads
\be
\int {\cal D}\phi {\cal D}\tilde{a} {\cal D}a \ \exp\Big(iS_{\rm scalar}[\phi;\tilde{a}]  + i S_{CS}[\tilde{a}] +  iS_{BF}[\tilde{a};a] + \frac{i}{2}S_{CS}[a] + \frac{i}{2}S_{BF}[a;A]\Big)\nn\ee
Integrating out $a$ results in the equation of motion $da= -(dA+2d\tilde{a})$. Substituting this back into the action and collecting terms, we find that
%
%
%
%
\be
\int {\cal D}\psi {\cal D}\tilde{a}  \ \exp\Big(iS_{\rm scalar}[\phi;\tilde{a}] - i S_{CS}[\tilde{a}] - iS_{BF}[\tilde{a};A] -  \frac{i}{2}S_{CS}[A] \Big)\nn\ee
As previously, we recognise the first three terms as the time reversed partition function $\bar{Z}_{\rm scalar+flux}[A]$ defined in \eqn{pscalarflux}. We replace this using the time reversed duality \eqn{pdual1}.  The upshot of this argument is that \eqn{dual3} implies the relationship between single flavour QED${}_3$, defined in \eqn{qed}, and a free fermion
\be Z_{\rm QED}[A] = Z_{\rm fermion}[A]\label{thisisgood}\ee
This is precisely the particle-vortex duality for fermions proposed in \cite{son,senthil,ashvin}, equating the partition functions for \eqn{son1} and \eqn{son2}.

\para
It is instructive to look at the quantum numbers of monopole operators in QED${}_3$ on ${\bf S}^2$. (See, for example, \cite{kapmono,ethan} for the necessary facts about monopole operators.) In the background of a monopole
with flux $\int da = 2\pi n$, the Dirac equation has $2|n|$ zero modes, transforming in the spin $(|n|-1)/2$ representation of the $SU(2)_{\rm rot}$ rotational symmetry.  For the $n=2$ monopole, the resulting states are $|0\rangle$, $\chi_a^\dagger |0\rangle$ and $\chi_1^\dagger\chi_2^\dagger |0\rangle$; these have charge $Q=-1,0,+1$ and spin $0,\frac{1}{2},0$ respectively. The Gauss law constraint projects us onto the $Q=0$ states. We learn that the monopole has spin $\frac{1}{2}$, as it should.

\para
The equality  of partition functions \eqn{thisisgood}, and the corresponding equality of current correlators, provides strong evidence that QED${}_3$ is indeed equivalent to a free Dirac cone. We stress that, on dynamical grounds, this is surprising. With an even number $N_f$ of fermionic flavours there is no parity anomaly and QED${}_3$ can be quantised with the usual flux condition \eqn{diracq}. Here the theory is expected to flow to a critical point when  $N_f > N_\star$, some critical number of flavours thought to be $N_\star\approx 4$. In contrast, for $N_f< N_\star$, the theory confines, and generates a gap spontaneously breaking the flavour symmetry. Based on this evidence, one might have thought that when $N_f=1$, the theory again confines and generates a gap, this time breaking time reversal invariance. However, the result \eqn{thisisgood} --- and, indeed, the arguments of \cite{senthil,ashvin} --- suggest that the theory does not confine. Presumably this is because the channel for time reversal breaking is somewhat weaker than flavour symmetry breaking  \cite{yesterday}.

\subsection{Self-Dual Theories}\label{sdsec}

It is straightforward to derive many further dualities by taking variations on this theme. Here we describe the self-dual theories.

\para
 A familiar story from the study of supersymmetric mirror symmetry is that when we couple two flavours of matter to a single $U(1)$ gauge field, the resulting theory is self-dual. This is known to hold for ${\cal N}=4$ \cite{intseib} and ${\cal N}=2$ \cite{doreytong,me} supersymmetric theories, which correspond to sigma-models with target space $T^\star{\bf CP}^1$ and ${\bf CP}^1$ respectively. In this section, we describe the non-supersymmetric analogs of these self-dual theories.

\subsubsection*{Self-Dual Fermions}

A proposal for a self-dual fermionic theory was offered recently in  \cite{cenke} by realising the theory on the surface of a topological insulator. Our derivation begins by putting together our original dual theory \eqn{dual1} with its time reversed partner \eqn{pdual1},
\be
Z_{\rm fermion}[A_1] Z_{\rm fermion}[A_2] = Z_{\rm scalar-flux}[A_1]
\bar{Z}_{\rm scalar-flux} [A_2] \, e^{+\frac{i}{2}S_{CS}[A_1] - \frac{i}{2}S_{CS}[A_2]} \label{dualhelp}\ee
We write the background gauge fields as
\be A_1 = a + C\ \ \ {\rm and}\ \ \ A_2 = a-C\nn\ee
We then promote $a$ to a dynamical gauge field, introducing a new background field $A$ in the process. The left-hand-side of the duality becomes
\be  Z_{{\rm QED}[N_f=2]}[A;C] = \int {\cal D}a\ Z_{\rm fermion}[a+C] Z_{\rm fermion}[a-C] \, e^{+iS_{BF}[a;A]} \nn\ee
The claim of \cite{cenke} is that this theory is actually self-dual in the sense that the physics is invariant under exchanging the two background fields $A$ and $C$.  This is not obvious from the expression above. Indeed, $C$ is the background field for the Cartan element of an $SU(2)$ flavour symmetry, rotating the two fermions. There is no obvious matching $SU(2)$ symmetry associated to $A$.

\para
We can use the duality \eqn{dualhelp} to help us. The right-hand side of \eqn{dualhelp} becomes
\be \int {\cal D}a {\cal D}\phi_1{\cal D}\tilde{a}_1{\cal D}\phi_2{\cal D}\tilde{a}_2 && \exp\Big(iS_{\rm scalar}[\phi_1;\tilde{a}_1] + iS_{\rm scalar}[\phi_2;\tilde{a}_2] +i S_{CS}[\tilde{a}_1]   - i S_{CS}[\tilde{a}_2] \nn\\ &&\ \ \ \ \ \ \ \ \  + \,i S_{BF}[\tilde{a}_1-\tilde{a}_2;a]  +  i S_{BF}[\tilde{a}_1+ \tilde{a}_2;C]+ i S_{BF}[a;A+C] \Big)\nn\ee
Once again, this doesn't look symmetric under interchange of $A$ and $C$. However now we can integrate out $a$. The equation of motion tells us that $d\tilde{a}_1 - d\tilde{a}_2 + dA + dC=0$. We will redefine $c_\pm = \tilde{a}_1 \pm \tilde{a}_2$ so that the constraint reads $dc_- = - (dA+dC)$ which we subsequently use to eliminate $c_-$.
 The kinetic terms for $\phi$ depend only on the symmetric combination $A+C$. Meanwhile, something rather nice happens to the remaining Chern-Simons and BF terms; they rearrange themselves so that they depend only on the combination $A-C$. We're left with
%
%
%
%
\be \int {\cal D}\phi_1{\cal D}\phi_2{\cal D}c_+\ \exp\Big(iS[\phi_1,\phi_2,c_+;A+C] - \frac{i}{2}S_{BF}[c_+; A-  C]  \Big)\nn\ee
We see that the first term is invariant under the exchange $A\leftrightarrow C$ while the second term picks up a minus sign. This, however, is easily dealt with if we simultaneously apply a time reversal transformation.

\para
Since this scalar theory is dual to QED with 2 flavours, we learn that this too must be self-dual under the interchange of $A\leftrightarrow C$, together with  time reversal
\be Z_{{\rm QED}[N_f=2]}[A;C] = \bar{Z}_{{\rm QED}[N_f=2]}[C;A]\nn\ee
in agreement with the proposal of \cite{cenke}.

\subsubsection*{Self-Dual Bosons}

It is a simple matter to repeat the steps above to derive the self-duality of $U(1)$ gauge field coupled to two scalars. Starting from the duality \eqn{dual7}, we find
\be Z_{{\rm QED}[N_s=2]}[A;C] &=& \int {\cal D}a \ Z_{\rm scalar}[a+C] Z_{\rm scalar}[a-C] e^{iS_{BF}[a;A]} \nn\\
&=& \int {\cal D}a\  Z_{\rm fermion+flux}[a+C]\, \bar{Z}_{\rm fermion+flux}[a-C] \, e^{ iS_{BF}[a;A-2C]} \nn\\
&=& \int {\cal D}\psi_1{\cal D}\psi_2 {\cal D}c_+ \ \exp\Big( i\tilde{S}[\psi_1,\psi_2,c_+;A-2C] - \frac{i}{4}S_{BF}[c_+;A+2C]\Big)
\nn\ee
where $\tilde{S}[\psi_1,\psi_2,c_+;A-2C]$ is what becomes of the kinetic terms after we integrate out $a$ and impose the resulting constraint $c_- = A-2C$. Importantly, this term is invariant under parity/time-reversal. We see that once again the partition function admits a symmetry under the exchange: $A\leftrightarrow -2C$ together with time reversal. We have the self-duality
\be Z_{{\rm QED}[N_s=2]}[A;C]  = \bar{Z}_{{\rm QED}[N_s=2]}[-2C;-\frac{1}{2}A] \nn\ee
This duality was previously studied in \cite{cp1}. Moreover, we learn something new: comparing the equations in this section, we see that QED with two fermions is actually the same theory as QED coupled to two scalars!

%
%
%
%
%
%
%

\subsection{A Vortex-Vortex Duality}

To finish, we describe one final duality in which monopole operators are mapped to monopole operators. This duality was previously described in \cite{ofer} for $U(N)$ theories; here we derive the $U(1)$ version.

\para
We again start with \eqn{dual2}, but this time change the Chern-Simons level for the background field on both sides to,
\be  Z_{\rm scalar}[C]\,e^{2iS_{CS}[C]} = Z_{\rm fermion+flux}[C]\,e^{iS_{SC}[C]} \nn\ee
After promoting $C$ to a dynamical gauge field (which we rename as $a$), the left hand side becomes
\be \int {\cal D}a\ Z_{\rm scalar}[a] \exp\Big( 2iS_{CS}[a] + iS_{BF}[a;A]\Big) \nn\ee
Meanwhile, the right-hand-side is
\be  && \int {\cal D}a{\cal D}C \ Z_{\rm fermi}[{a}]\,\exp\Big(
-\frac{i}{2}S_{CS}[{a}] - iS_{BF}[a;C] + iS_{CS}[C] +iS_{BF}[C;A]\Big)  \nn\\ &=&  \int {\cal D}a \ Z_{\rm fermi}[a]\,\exp\Big(-\frac{3i}{2}S_{CS}[a] +i S_{BF}[a;A] - S_{CS}[A]\Big)\nn\ee
where, to get to the second line, we integrate out $C$ and substitute in its equation of motion $C=a-A$. The end result is that a scalar coupled to a Chern-Simons gauge field at level 2 is equivalent to a fermion coupled to a Chern-Simons gauge field at level $-\frac{3}{2}$. This duality was previously reported in \cite{ofer}.

\para
Let us check that the quantum numbers of operators agree on both sides. On the scalar side, Gauss' law requires $\rho_{\rm scalar} = - f/\pi$. In the background of a single monopole, we must turn on two scalar modes. As we saw above, the lowest excited state of the scalar has spin $\frac{1}{2}$. Since these are bosons, we pick out the symmetric part so turning on two such modes endows the monopole with spin $1$.

\para
Meanwhile, on the fermionic side the Gauss' law tells us that $\rho_{\rm fermi} = 3 f/4\pi$. The single monopole must have charge $3/2$. The zero mode \eqn{twostates} can account for charge $\frac{1}{2}$ and leaves the monopole with spin $0$. But, in addition, we must also turn on an excited mode. The first excited mode has spin $1$. Thus, once again, the monopole operator has spin $1$.

\section{Discussion}\label{dissec}

Above, we have focused exclusively on non-supersymmetric dualities. One may wonder if we can combine these to derive mirror symmetry \cite{intseib} or Seiberg-like dualities \cite{givkut} of supersymmetric theories. Unfortunately we do not, at present, have enough handle on the operator map needed to include features like Yukawa couplings on both sides.

\para
Instead, we could ask the reverse question: starting from a supersymmetric mirror pair, how does it decompose under RG flow if a relevant, supersymmetry breaking operator is added? For the supersymmetric bosonization duality, this question was answered in the large $N$ limit in \cite{jain,guy3}. Here our interest is in Abelian theories and we do not have control of the RG flow. Nonetheless, we will make some suggestions.

\para
 For theories with ${\cal N}=4$ supersymmetry, the simplest mirror pair is \cite{intseib}
\be \mbox{ Free Hypermultiplet = $U(1)$ + Charged Hypermultiplet }\nn\ee
Here a single hypermultiplet contains two complex scalars and two Dirac fermions. The $U(1)$ gauge field is part of a vector multiplet which also contains three real scalars and two Dirac fermions. It can be shown that, starting from this seed mirror pair, one can generate all further ${\cal N}=4$ Abelian mirrors using path integral manipulations of the kind employed above \cite{kapstrass}.

\para
One can break the supersymmetry down to ${\cal N}=2$ by giving masses to half of the fields in the hypermultiplet and integrating them out. Flowing from the  simple ${\cal N}=4$ duality above, one finds  the ${\cal N}=2$ duality involving a half-integer Chern-Simons coupling  \cite{doreytong,me}
\be \mbox{Free Chiral Multiplet = $U(1)_{1/2}$ + Charged Chiral Multiplet}\nn\ee
Now the relevant ${\cal N}=2$ multiplets are a chiral multiplet, which consists of a single complex scalar and a single Dirac fermion, and a vector multiplet containing the $U(1)$ gauge field, a single real scalar, and a single Dirac fermion.

\para
Now we wish to deform the theory further by adding mass to either the free boson or free fermion. The resulting RG flow breaks all supersymmetry and is, correspondingly, difficult to study. One might imagine that the end point of integrating out the fermions  on both sides is the standard bosonic particle-vortex duality \eqn{xy} and \eqn{ah}. However, the end point of integrating out scalars on both sides cannot be the fermionic counterpart \eqn{son1} and \eqn{son2} because the anomaly structure is different. Something else must happen.

\para
Flows of a very similar kind were studied in a very impressive analysis of  large $N$ theories in \cite{jain,guy3}. There one finds that the bosonic currents on one side of the duality are related to the fermionic currents on the other side. Further, gapping out the  boson on one side ultimately results in a gap for fermion on the other side. We suggest that this structure survives to the Abelian theories considered here: the end points of  RG flows from the ${\cal N}=2$ mirror are not the particle-vortex dualities described in the introduction: instead they are the two bosonization dualities  \eqn{dual1} and \eqn{dual2}.

\para
The results of Section \ref{pvsec}, as well as our speculations above, fit a general pattern which suggests that a good slogan for the content of this paper might be: ``particle-vortex duality  = bosonization$^2$\,".

\para
Our techniques open the door for the construction of many more Abelian dual pairs. In the supersymmetric context, one can use $N$ copies of the basic pair and gauge $r$ of the background fields to derive mirror pairs in which $U(1)^r$ with $N$ matter fields is dual to $U(1)^{N-r}$ with $N$ matter fields  \cite{intseib,kapstrass,me}. The same steps can easily be repeated in the non-supersymmetric context to generate generalizations of our story with multiple Abelian gauge groups, with the self-dual theories of Section \ref{sdsec} the first in the sequence.

\para
More challenging is the extension of our techniques to non-Abelian theories. The 3d bosonization dualities have a simple non-Abelian generalization. Indeed, in the large $N$ limit the evidence for this duality is overwhelming. It would be very interesting to see if one can manipulate the partition functions for these non-Abelian duals in a similar way to their Abelian cousins.

\section*{Acknowledgements}

We are supported by the US Department of Energy under grant number DE-SC0011637, by STFC, and by the European Research Council under the European Union's Seventh Framework Programme (FP7/2007-2013), ERC grant agreement STG 279943, ``Strongly Coupled Systems". DT is grateful to the Stanford Institute for Theoretical Physics and to the University of Washington for their kind hospitality while this work was undertaken. Our thanks to Ofer Aharony, Ethan Dyer, Andrew Essin,  Yin-Chen He, Guy Gur-Ari, Shamit Kachru, Max Metlitski, Shiraz Minwalla, Mike Mulligan, Sri Raghu, Eva Silverstein, Dam Son,  Senthil Todadri, Christoph Uhlemann and  Ashvin Vishwanath for many illuminating comments and conversations.  Our thanks to Nati Seiberg, Senthil Todadri, Chong Wang and Edward Witten for communicating their results prior to publication.

\end{document}